\documentclass[letter]{aa}  
\usepackage{graphicx}
\usepackage{txfonts}

\usepackage{natbib}
\begin{document}
\title{Discovery of a magnetic field in the CoRoT hybrid B-type pulsator HD~43317\thanks{Based on observations obtained using the Narval spectropolarimeter at the Observatoire du Pic du Midi (France), which is operated by the Institut National des Sciences de l'Univers (INSU).}}


   \author{M. Briquet\inst{1}\fnmsep\thanks{F.R.S.-FNRS Postdoctoral Researcher, Belgium}
          \and
          C.  Neiner\inst{2}
          \and
          B. Leroy\inst{2}
          \and
          P.I. P\'apics\inst{3}     
          \and
          the MiMeS collaboration}
   \institute{Institut d'Astrophysique et de G\'eophysique, Universit\'e de Li\`ege, All\'ee du 6 Ao\^ut 17, B\^at B5c, 4000 Li\`ege, Belgium\\
              \email{maryline.briquet@ulg.ac.be}
         \and
         LESIA, Observatoire de Paris, CNRS UMR 8109, UPMC, Univ. Paris Diderot, 5 place Jules Janssen, 92195 Meudon Cedex, France        
           \and  
         Instituut voor Sterrenkunde, Katholieke Universiteit Leuven, Celestijnenlaan 200 D, 3001 Leuven, Belgium
 }

   \date{Received ; accepted}

 
  \abstract
   {A promising way of testing the impact of a magnetic field on internal mixing (core overshooting, internal rotation) in main-sequence B-type stars is to perform asteroseismic studies of a sample of magnetic pulsators.}
    {The CoRoT satellite revealed that the B3IV star HD~43317 is a hybrid SPB/$\beta$~Cep-type pulsator that has a wealth of pulsational constraints on which one can perform a seismic modelling, in particular, probing the extent of its convective core and mixing processes. Moreover, indirect indicators of a magnetic field in the star were observed: rotational modulation due to chemical or temperature spots and X-ray emission. Our goal was to directly investigate the field in HD~43317 and, if it is magnetic, to characterise it.} 
    {We collected data with the Narval spectropolarimeter installed at TBL (T\'elescope Bernard Lyot, Pic du Midi, France) and applied the least-squares deconvolution technique to measure the circular polarisation of the light emitted from HD~43317. We modelled the longitudinal field measurements directly with a dipole.}
   {Zeeman signatures in the Stokes V profiles of HD~43317 are clearly detected and rotationally modulated, which proves that this star exhibits an oblique magnetic field. The modulation with the rotation period deduced from the CoRoT light curve is also confirmed, and we found a field strength at the poles of about 1 kG. Our result must be taken into account in future seismic modelling work of this star.}
    {}

   \keywords{stars: magnetic field -- stars: individual: HD~43317}

   \maketitle
%

\section{Introduction}
The study of the magnetic properties of pulsating B-type stars - $\beta$~Cep and slowly pulsating B (SPB) stars - is particularly interesting because when it is combined with the study of their pulsational properties, it provides a unique way of probing the impact of magnetism on the physics of non-standard mixing processes inside these stars. Comparing the amount of mixing obtained by asteroseismic investigation for a sample of magnetic $\beta$~Cep stars with that of a sample of non-magnetic objects would already allow us to corroborate or disprove that magnetic fields inhibit mixing in stellar interiors. Moreover, the surface field strength obtained from modelling the magnetic measurements is an additional constraint for a comparison with the internal field strength needed to inhibit mixing in the radiative zone, as obtained by different criteria (e.g., \citealp{mathis2005}).

Up to now, asteroseismic modelling has been performed for two magnetic $\beta$~Cep stars only: $\beta$~Cephei itself \citep{shibahashi2000} and V2052~Oph \citep{briquet2012}. For V1449 Aql, which was also modelled asteroseismically \citep{aerts2011}, a magnetic field was detected by \cite{hubrig2011}, but \cite{schultz2012} disputed the presence of this field. Magnetic fields have also been detected in a number of other $\beta$~Cep and SPB stars (e.g., \citealp{donati2001}, \citealp{hubrig2006}, \citealp{hubrig2009}), but there is no seismic modelling available for them. Moreover, the approach presented in the previous paragraph has been accomplished only for V2052~Oph so far \citep{briquet2012,neiner2012a}. Asteroseismic investigations of this magnetic target, which has a polar field strength of about 400 G \citep{neiner2012a}, showed that the stellar models that explained the observed pulsational behaviour needed no convective core overshooting \citep{briquet2012}. This outcome is in contrast to other results of dedicated asteroseismic studies of non-magnetic $\beta$~Cep stars (e.g., \citealp{briquet2007}): it is usually found that convective core overshooting needs to be included in the stellar models \citep{aerts2010}. The most plausible explanation is that the magnetic field inhibits non-standard mixing processes inside V2052~Oph. Indeed, the field strength observed in this star is higher than the critical field limit needed to inhibit mixing that was determined from theory \citep[see][]{zahn2011,neiner2012a}. 

A recent study of the B3IV single star HD~43317 revealed it to be a promising target to explore the effects of magnetism on stellar interiors of main-sequence B-type stars in more detail. Indeed, an analysis of the CoRoT light curve of this star showed $p$ and $g$ pulsation modes of $\beta$~Cep and SPB types, as well as two series of consecutive frequencies with an almost constant period spacing near 6300 seconds \citep[][hereafter P12]{papics2012}. Such an observation carries information on the extent of the convective region and mixing processes \citep{miglio2008}, as applied to only one hybrid B-type pulsator so far \citep{degroote2010}. Moreover, if the star is magnetic, the detected high-order g-modes have the potential to probe its internal magnetic field, as suggested by the theoretical work of \cite{hasan2005}; but this has never been applied to a massive star.

In both the photometry and spectroscopy of P12, rotational modulation connected to chemical or temperature spots were observed, providing us with a precise value of the rotation period of the star ($P_{\rm rot}$=0.8969 days), which rotates at 50\% of its critical velocity. The clear signature of rotational modulation in the He lines, as shown and discussed in P12 (see their Fig.\,7), is an indirect indicator of a magnetic field in HD~43317. Indeed, it is generally believed that magnetic fields are responsible for the formation of chemical spots at the surface of chemically peculiar Bp stars (e.g., {\citealp{michaud1981}). Moreover, X-ray emission for our object is reported in the ROSAT all-sky survey catalogue of optically bright OB-type stars   with $L_{\rm X}$ $\sim$ 8.3 x 10$^{30}$ erg~s$^{-1}$ \citep{berghoefer1996}. This may also point towards a magnetic field, because X-rays can be produced by shock waves in the stellar winds or by confinement of matter at the magnetic equator. 

In this paper, we present the first direct detection of a magnetic field in HD~43317 by making use of the high-efficiency and high-resolution Narval spectropolarimeter installed at the TBL (T\'elescope Bernard Lyot) 2-m telescope (Pic du Midi, France). 


\section{Observations}\label{obs}

Nineteen high-resolution (R=65000) spectropolarimetric Stokes V Narval observations of HD~43317 were collected in 2012 (see Table\,\ref{log}). The first four measurements (of 4*1000s each) were obtained to investigate the presence of a magnetic field. A Zeeman signature being detected, we continued our sequence of observations to sample other rotational phases, but changed our observing strategy to consider the fact that the star is pulsating and to increase the signal-to-noise ratio (S/N). The next 15 observations consisted of 5 $\times$ 3 successive sequences (which were afterwards averaged) of 4*800s each. Each sequence was shorter than 1/20th of the pulsation period to avoid that the deformations of the intensity profile by pulsations during the measurement polluted the Stokes V signal. Because the main pulsation period detected in the spectroscopy of HD~43317 is 0.735~d (P12), the exposure time is expected not be longer than 4x800s at once. 

\begin{table}
\centering
\caption{Journal of Narval/TBL observations and magnetic field measurements of HD~43317.}
\begin{tabular}{ccccccc}
\hline\hline
Nr. &    Mid-HJD & $T_{\rm exp}$ & Phase  &  $B_l$ &  S/N \\
    &    $-$2456000 & s & & G &  \\
\hline
1& 185.66197 & 4x1000 & 0.35710& $-$64.9$\pm$47.1 &  3860\\
2& 203.61791 & 4x1000 & 0.37792& $-$48.1$\pm$34.3 &  5299\\
3& 206.64675 & 4x1000 & 0.75506& 55.6$\pm$36.7  &  4901\\
4& 214.58370 & 4x1000 & 0.60474& $-$141.1$\pm$44.9 &  3971\\
5& 230.60774 &  3x(4x800)& 0.47150& $-$79.3$\pm$22.4  &  7962\\  
6& 232.58893 &  3x(4x800)& 0.68052& $-$90.8$\pm$25.9  &  6862 \\ 
7& 244.60017 &  3x(4x800)& 0.07301& 176.8$\pm$30.9  &  5756 \\     
8& 245.67528 &  3x(4x800)& 0.27176& 124.6$\pm$50.6  &  3595\\	   
9& 254.55404 &  3x(4x800)& 0.17155& 151.1$\pm$45.8  &  3939\\      
\hline
\end{tabular}
\tablefoot{Column~1 indicates the number of the magnetic measurement. Column~2 gives the heliocentric Julian date (HJD) at the middle of the observations, and Col.~3 gives the total exposure time in seconds. Column~4 provides the rotational phase using $P_{\rm rot}$=0.8968634~d and the reference date HJD$_0$=2455174.57665. The longitudinal magnetic field value $B_l$ in Gauss extracted from LSD profiles is given in Col.~5, and the S/N ratio per 2.6 km~s$^{-1}$ pixels in the LSD Stokes V profile is indicated in Col.~6.}
\label{log}
\end{table}

The usual bias, flat-fields, and ThAr calibrations were obtained each night. The data reduction was performed using {\sc libre-esprit} \citep{donati1997}, the dedicated software available at TBL. The intensity spectra were then normalised to the continuum by fitting a cubic spline function. Stokes V spectra were derived by constructively combining the left- and right-hand circularly polarised light together from the four sub-exposures. The three measurements obtained successively each night were averaged so that we finally had nine magnetic measurements that covered different phases of rotation. They were normalised to the continuum intensity.

We adopted the ephemeris of HD~43317 from P12, that is, a rotation period
$P_{\rm rot}=$0.8968634~d and a  reference date HJD$_0$=2455174.57665.

We applied the least-squares deconvolution (LSD) technique \citep{donati1997} to the photospheric spectral lines in each \'echelle spectrum ($\lambda$=[3750--10500] \AA) to construct a single profile with an increased S/N (note that no additional normalisation was applied to the LSD profiles). In that way, the Zeeman signature induced by a magnetic field is clearly visible in the high S/N Stokes V profiles (Fig.\,\ref{stokes}), proving that HD~43317 is a magnetic object. To diagnose possible spurious polarisation signatures, the null profiles N were also computed by destructively combining the four subexposures. The flatness of these profiles (Fig.\,\ref{stokes}) indicates that none of our measurements has been polluted by stellar pulsations, observing conditions, or instrumental effects, even those with slightly longer individual exposure times. To compute the LSD profiles, we made use of a mask created from the Kurucz atomic database and ATLAS~9 atmospheric models of solar abundance \citep{kurucz1993}, for T$_{\rm eff}$=17000 K and $\log g$=4.0 (following P12), with intrinsic line depths higher than 0.1. This mask contains 199 photospheric He and metal lines of various chemical elements, together with their wavelength, depth, and Land\'e factor. The depths were modified so that they correspond to those of the observed spectral lines, with an interactive graphical user interface (GUI) program, written in IDL, that was kindly put at our disposal by J.H. Grunhut. The adjustment was straightforward for unblended individual lines. The relative depths of the blended lines in the original mask were kept fixed and the entire line blend depth was adjusted to the spectrum. Lines whose Land\'e factor is unknown or that did not appear in the spectrum were excluded. The average S/N is 5100 per 2.6 km~s$^{-1}$ pixels in the LSD Stokes V profiles.

\begin{figure}
\begin{center}
\includegraphics[width=\hsize,clip]{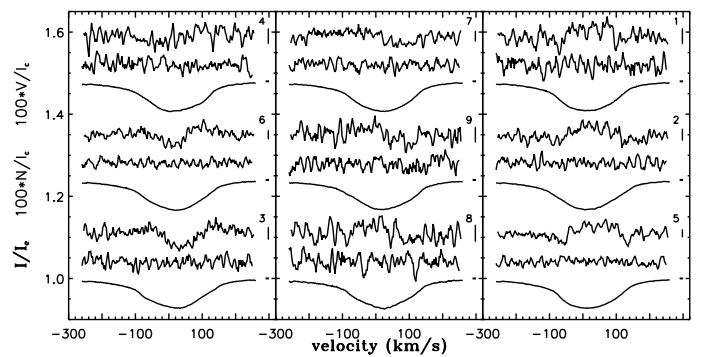}
\caption{ LSD Stokes V (top), N (middle), and I (bottom) profiles for the nine rotational phases. The number of the magnetic measurement, as listed in Table\,\ref{log}, is indicated on the right of each profile. Typical error bars are shown next to each I and V profile.
} 
\label{stokes}
\end{center}
\end{figure}

\section{Magnetic field}\label{mag}

\subsection{Longitudinal magnetic field}\label{sect_Bl}

We used the LSD Stokes I and V profiles to compute the line-of-sight component of the magnetic field integrated over the visible stellar surface, that is, the longitudinal magnetic field in Gauss, given for instance by Eq.~(1) of \cite{wade2000}. In this equation we used $\lambda$=501 nm and g=1.20. The integration limits were chosen to be high enough to cover the line width but also small enough to avoid artificially increased error bars due to noisy continuum. A range of 123 km~s$^{-1}$ around the line centre was adopted.

The values for the longitudinal magnetic field are reported in Table\,\ref{log}. They vary between about $-$140 and 180 G.  The error bars are typically $\sim$35 G. A phase diagram of the longitudinal field folded with the rotation period is shown in Fig.\,\ref{blfigure}. 

From a dipolar fit of the $B_l$ measurements, we derived $B_{\rm l,max}=193 \pm 30$ G, $B_{\rm l,min}=-115 \pm 30$ G, and the phase difference $\Delta\Phi=0.58 \pm 0.02$ between two successive crossings of the $B_l$ fit at 0. The dipole fit has a reduced $\chi^2$ value of 1.3. From Eqs.~(18) and (19) of \cite{shore1987}, for instance, we then derived the obliquity angle $\beta$ $\in$ [70, 86] deg, assuming the inclination angle $i$ $\in$ [20, 50] deg as deduced for the star by P12, but considering the errors on the $T_{\rm eff}$ and $\log g$ values.

With these values of $i$ and $\beta$, we also deduced the polar field strength $B_{\rm pol}$ $\in$ [650, 1750] G from Eq.~(5) of \cite{borra1979}.

\begin{figure}
\begin{center}
\includegraphics[width=1\hsize,clip]{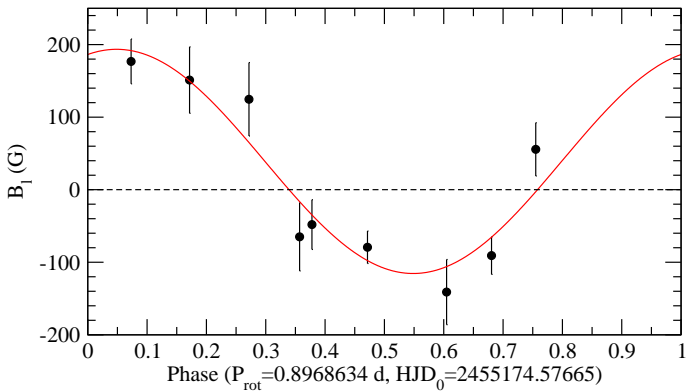}
\caption{Longitudinal magnetic field curve of HD~43317, folded in phase with the rotation period $P_{\rm rot}$=0.8968634~d and a reference date HJD$_0$=2455174.57665 from P12. A sinusoidal (dipole) fit is plotted in red.}
\label{blfigure}
\end{center}
\end{figure}

\subsection{Magnetic confinement and magnetosphere}

We checked whether the magnetic field we discovered in HD~43317 is sufficient to confine the stellar wind. We used the magnetic confinement parameter, $\eta_*$, defined by \cite{uddoula2002}, which depends on the stellar equatorial magnetic field strength, radius, and wind momentum. When $\eta_* > 1$, wind material can be channelled along the field lines and be confined into a circumstellar magnetosphere.

To be conservative, we used the lower value of the polar field strength of $B_{\rm pol}$=650 G discovered in this work and an upper value of the mass loss of $\dot{M}$=$10^{-9}$ M$_\odot$ yr$^{-1}$ according to Fig.~4 of \cite{petit2013} for a B3IV star. We measured the terminal wind velocity $v_{\infty} \sim 150$ km~s$^{-1}$ from the \ion{Si}{iv} 1394+1403 \AA\ and \ion{Al}{iii} 1855+1863 \AA\ lines available in the SWP IUE archival spectrum. Finally, we used $i$ $\in$ [20, 50] deg as derived above. We imposed that $P_{\rm rot}\sim0.89$ d. We found that $\eta_*$=3400 for $i$=50 deg and $\eta_*$=14500 for $i$=20 deg. In addition, we obtained that the Alfv\'en radius is $R_A$=7.6 R$_*$ for $i$=50 deg and $R_A$=11 R$_*$ for $i$=20 deg, and the corotating Kepler radius is $R_K$=2.6 R$_*$ for $i$=50 deg and $R_K$=1.2 R$_*$ for $i$=20 deg, respectively. Therefore, wind material is confined by the magnetic field ($\eta_* > 1$) and $R_A > R_K$, that is, the magnetosphere is supported by the centrifugal force. See \cite{uddoula2008}, \cite{petit2012} and \cite{neiner2012b} for more details on magnetospheres.

We did not detect emission in the H$\alpha$ line of HD~43317. This means that the emission measure in the centrifugally supported magnetosphere is too low, probably because the wind is weak and rather slow. Nevertheless, we recall that HD~43317 is detected as an X-ray source in the ROSAT catalogue \citep{berghoefer1996} with $\log (L_X/L_{\rm bol})$=$-5.94$ and an estimated X-ray temperature of 0.49 keV. X-ray emission could be produced in the magnetic equator where wind particles from the two magnetic hemispheres collide.

\section{Conclusions}\label{sum}

Thanks to photometric data of unprecedented precision assembled by the CoRoT satellite complemented by ground-based spectroscopy (P12) and spectropolarimetry (presented in this paper), HD~43317 was discovered to be a single magnetic B-type hybrid pulsator with much observational information that can be used to constrain its interior properties and to study the effect of magnetism in the mixing processes that act inside the star. In future work on HD~43317, the magnetic field must be taken into account, for example to compute the oscillation periods that need to be compared with the two almost constant period spacings of high-order gravity modes observed in the star (Salmon et al., in preparation).

HD~43317 is also an interesting target in which to study physical mechanisms that take place at its stellar surface, such as the interplay between radiatively driven diffusion and magnetic field, which is expected to be responsible for the surface abundance inhomogeneities. To this end, additional spectropolarimetric data that cover the rotation cycle are required so that a Zeeman-Doppler imaging of the stellar surface can be performed, which will provide a detailed characterisation of the magnetic configuration as well as abundance mapping of the stellar surface. 

\begin{acknowledgements}
We thank Evelyne Alecian for fruitful discussions. The research leading to these results has received funding from the European Research Council under the European Community's Seventh Framework Programme (FP7/2007--2013)/ERC grant agreement n$^\circ$227224 (PROSPERITY), as well as from the Belgian Science Policy Office (Belspo, C90309: CoRoT Data Exploitation).
\end{acknowledgements}
\bibliographystyle{aa}
\bibliography{hd43317_MB}

\end{document}